\documentclass[a4,12pt]{elsarticle}
\usepackage{graphicx}
\usepackage{amsmath}
\usepackage{amssymb}
\usepackage{amsthm}
\begin{document}
\begin{frontmatter}
\title{Nonlinear optical properties in a quantum well with the hyperbolic confinement potential}
\author{Tairong Chen$^{\ast}$ \footnote{E-mail: Chen-tairong4487@hotmail.com}\quad Wenfang Xie \quad  Shijun Liang}

\address{(Department of Physics, College of Physics and
Electronic Engineering, Guangzhou University, Guangzhou 510006, P.R. China)}
\begin{abstract}
We have performed theoretical calculation of the nonlinear optical properties in
a quantum well (QW) with the hyperbolic confinement potential. Calculation results reveal that the transition energy, oscillator strength, second-order nonlinear optical rectification (OR), geometric factor and nonlinear optical absorption (OA) are strongly affected by the parameters ($\alpha, \sigma$) of the hyperbolic confinement potential. And an increment of the parameter $\alpha$ reduces all these physical quantities, while an increment of the parameter $\sigma$ enhances them, but not for geometric factor. In addition, it is found that one can control the optical properties of QW by tuning these parameters.
\end{abstract}
\end{frontmatter}
\section{\textbf{Introduction}}
Nonlinear optical properties of low-dimensional semiconductor nanostructures such as quantum wells (QWs), quantum wires, and quantum dots have become a very active field of research with respect to both theory and experiment [1-5] in the past decades. The advance in nanofabrication techniques has made it possible for researchers to fabricate ultra-small
semiconductor structures of low nanometer size [6,7], which has an important role and displays an interesting behavior in microelectronic and optoelectronic devices due to their influences on the electrical, optical and transport properties. And a number of theoretical and experimental investigations have been published.

It is well known that the pseudopotential plays an important role in the studies of semiconductor structures [8-14]. Nowadays, more and more important information about electronic structures in semiconductor have been theoretically obtained by using the pseudopotential approaches. Tsirkin $et$ $al.$ [8] studied the model pseudopotential for the $ (1 1 0)$ surface of face-centered cubic (fcc) noble metals. The result shows that the proposed two-dimensional model pseudopotential allows to accurately reproduce the electronic structure of the $ (1 1 0)$ surface of fcc noble metals at $\bar{Y}$ point. Wang $ et$ $al.$ [9] reported the pseudopotential calculation of nanoscale CdSc quantum dots, and Leiniger $et$ $al.$ [10] investigated the
accuracy of the pseudopotential approximation. $\uppercase\expandafter{\romannumeral 2 }$. A comparison of various core sizes for indium pseudopotentials in calculations for spectroscopic constants of $InH$, $InF$, and $InCl$. Moreover, the pseudopotential is applied to interpreting some results from experiments with great success [15,16].

We have paid great attention to the parabolic potential before, which, as one of the pseudopotentials, is often considered to be a good representation of the potential in a quantum dot. The effects of the parabolic potential on the energy and optical properties have been widely investigated. Recently, the theory of boundary value problems for hyperbolic partial differential
equations has been clarified to a large extent. With separation of
variable techniques, Ladyzhenskaya $et$ $al.$ [17] investigated  the general
theorems on the existence of classical solutions to hyperbolic equations with
coefficients depending on the spatial variables. Lu $et$ $al.$ [18] studied the rotation and vibration of diatomic molecule
oscillator with hyperbolic potential function. To our knowledge, there has been no
report of the nonlinear optical properties in a QW with the hyperbolic confinement potential. With respect to the lack of such studies, we believe that the present study can make an important contribution to the literatures.
The aim of this paper is to investigate the effects of the hyperbolic potential on the nonlinear optical properties in a quantum well. For this purpose, we focus on studying the transition energy, oscillator strength, geometric factor and second-order nonlinear optical rectification (OR) coefficient. The paper is organized as follows:
The model and theoretical framework is described in section 2, the results and discussions are presented in section 3, and finally, the conclusions are given in section 4.
\section{\textbf{Model and theory}}
\subsection{Electronic state in QWs with the  hyperbolic confinement potential }
Consider an electron in the hyperbolic potential well, X axis is perpendicular to the well layers, so Hamiltonian within the framework of the effective mass approach is given by
\begin{equation}
H=\frac{p_{x}^2}{2m_{e}}+V(x),
\end{equation}
where $m_{e}$ is the electronic effective mass. $V(x)$ is the hyperbolic  potential which was proposed by Schi\"{o}berg as [19]
\begin{equation}
V(x)=D[1-\sigma\coth{(\alpha x)}],
\end{equation}
where $D$, $\sigma$ and $\alpha$ are three parameters representing the potential property. Fig. 1 shows the shape of the hyperbolic potential well for the different parameter $\alpha$ with $\sigma=0.02$ and the different parameter $\sigma$ with $\alpha=2.0\times10^{8}radm^{-1}$, respectively. From Fig. 1(a) one can see that, at the beginning, the values of $V/D$ decrease rapidly with increasing $x$, and then they reach a minimum at around some $x$. As $x$ is further increased, the values begin to increase, and finally, they all approach asymptotically unity. Moreover, the
minimum position shifts to lower $x$ as $\alpha$ increases. On the other hand, from Fig. 1(b), one can find that the shape of the hyperbolic potential well is similar to that from Fig. 1(a), but the minimum position shifts to higher $x$ as $\sigma$ increases.
The schr\"{o}dinger equation with this potential is written as  [20]
\begin{equation}
-\frac{\hbar^2}{2m_{e}}\frac{d^2}{d x^2}\psi(x)+D[\sigma^2 csch^2{(\alpha x)}-2
\sigma\coth{(\alpha x)}]\psi(x)=\tilde{E}\psi(x),
\end{equation}
where we use the following notation
\begin{equation}
\tilde{E}=E-D(1+\sigma^2),
\end{equation}
and the formula
\begin{equation}
csch^2{(\alpha x)}=\coth^2{(\alpha x)}-1.
\end{equation}

The wave function and the energy spectrum of an electron confined in hyperbolic potential can be obtained [20] by solving the Schr\"{o}dinger equation.
\begin{equation}
\psi(z)=N(1-z)^{\delta+1}z^{\beta}{}_{2}F_{1}(-n,n+2(\delta+\beta+1);2\beta+1;z),
\end{equation}
and
\begin{equation}
E_{n}=D(1+\sigma^2)-[\frac{m_{e}U_{1}^2}{2\hbar^2(C+n \alpha)^2}+\frac{\hbar^2(C+n \alpha)^2}{2m_{e}}],
\end{equation}
where the new parameters are defined as follows [20]: $U_{1}=2D\sigma$, $z=e^{-2\alpha x}$, $k=\frac{m_{e}D}{2\alpha^2\hbar^2}$, $\delta=\frac{1}{2}[-1+\sqrt{1+16k \sigma^2}]$, $\lambda=\sqrt{\frac{-m_{e}E}{2\alpha^2\hbar^2}} $ and $ \beta=\sqrt{\lambda^2+k(1-\sigma)^2}$. And $N$ is the normalized factor to be determined from the normalization condition $\sum \psi_{z}^2dx=1$, from which $N$ can be further obtained as [20]
\begin{equation}
N=\sqrt{\frac{4\alpha \beta(1+2\beta)_{n}\Gamma(n+2(\delta+2\beta+2))}
{2n!(n+\delta+1)(n+2(\delta+\beta+1))_{n}\Gamma(1+2\beta)\Gamma(n+2\delta+2)}},
\end{equation}
where the Pochhammer symbol $(a)_{n}$ is defined as [20,21]
\begin{equation}
(-a)_{n}=\frac{(-1)^n a!}{(a-n)!}.
\end{equation}
\subsection{Oscillator strength }
The oscillator strength is a very important physical quantity in the study of the optical properties, which are related to the electronic dipole-allowed transition. Furthermore, the oscillator strength can offer additional information on the fine structure and selection rules of the optical absorption. Generally, the oscillator strength $P_{fi}$ is defined as
\begin{equation}
P_{fi}=\frac{2m_{e}}{\hbar^2}\triangle E_{fi}|\mu_{fi}|^2,
\end{equation}
where $\Delta E_{fi}=E_{f}-E_{i}$ donates difference of the energy between lower and upper level, $m_{e}$ is the electronic effective mass, and $\mu_{fi}=<\psi_{i}|x|\psi_{f}>$ is the electric dipole moment of the transition from the $\psi_{i}$ state to the $\psi_{f}$ state.

\subsection{Nonlinear optical rectification and nonlinear optical absorption in hyperbolic potential QWs }
Based on the density matrix approach and perturbation expansion method, the second-order nonlinear optical rectification coefficient is given by [22-25]
\begin{equation}
\chi_{0}^{(2)}=\frac{4e^3\rho \mu_{fi}^2\delta_{fi}\{\Delta E_{fi}^2(1+\frac{T_{1}}{T_{2}})
+[(\hbar \omega)^2+(\frac{\hbar}{T_{2}})^2](\frac{T_{1}}{T_{2}}-1)
\}}{\epsilon_{0}[(\Delta E_{fi}-\hbar\omega)^2+(\frac{\hbar}{T_{2}})^2][(\Delta E_{fi}+\hbar\omega)^2+(\frac{\hbar}{T_{2}})^2]},
\end{equation}
additionally, in the present work, we also investigate the intensity-dependent resonant peak of the nonlinear optical absorption coefficient which is given by [26]
\begin{equation}
\eta_{max}=\eta (\omega=\omega_{fi};I=0) =\frac{e^2\rho \Delta E_{fi}\mu_{fi}^2T_{2}}{\hbar^2\varepsilon_{0}c n_{r}},
\end{equation}
where $\varepsilon_{0}$ is the vacuum permittivity, $\rho$ donates the volume density of electrons in the structure. $e$ is the electronic charge. $T_{1}$ and $T_{2}$ are the longitudinal and the transverse relaxation times, respectively. In calculation we set the relaxation time at $T_{1}=1 ps$, $T_{2}=0.2 ps$ [24,25]. And $\delta_{fi}=|\mu_{ff}-\mu_{ii}|=|<\psi_{f}|x|\psi_{f}>-<\psi_{i}|x|\psi_{i}>|$ [24], $\hbar\omega$ is the photon energy. $c$ is the vacuum velocity of light, and $n_{r}$ is the refractive index of the semiconductor.

\section{\textbf{Result and Discussions}}
We carry out numerical calculations for a typical  $GaAs/AlGaAs$ QW with the hyperbolic confinement potential. The parameters chosen in the present study are the followings [24,25]: $m_{e}=0.067m_{0}$, where $m_{0}$ is the free electron mass, $\rho=5\times10^{24}m^{-3}$, and $D=0.5\times10^{5}meV$. We restrict our study to the transition between $\psi_{0}$ state and $\psi_{1}$ state.

The results of our calculations are presented in Figures 2-6. In Fig. 2,
we display the theoretical results for the transition  energy in a hyperbolic potential QW as a
function of $\alpha$ for three different values of $\sigma$. From Fig. 2, it is clearly found that the transition energy decreases monotonously as $\alpha$ increases. The physical origin is that the increment of $\alpha$ results in a reduction of the confinement strength of the hyperbolic potential QWs. On the other hand, one can see that the transition energy increases with the increase of $\sigma$. This is because the confinement strength enhances as $\sigma$ increases, so one can  conclude that
the two parameters $(\alpha, \sigma)$ can greatly affect the energy level distribution of the electronic states in a hyperbolic potential QW, moreover,
the electron wave function is more strongly localized inside the hyperbolic QW with decreasing $\alpha$ and increasing $\sigma$, respectively.

Fig. 3 presents the variation of the oscillator strength as a function of $\alpha$ for different $\sigma$. Obviously, one can observe that the oscillator strength decreases as $\alpha$ increases, but finally, they all approach asymptotically zero as $\alpha$ increases further. The physical explanation is that the increasing $\alpha$ leads to a reduction of
the energy difference between the two different electronic states, as well as the electric dipole moment. Hence, the tunneling probability for hyperbolic potential becomes smaller. Moreover, it is shown that the oscillator strength increases with  increment of $\sigma$. The main reason is that the increasing $\sigma$ causes an enhancement of the energy difference between the initial state and the final state. and there is a bigger tunneling probability with the increasing $\sigma$, as a result, the oscillator strength enhances.

Fig. 4(a), 4(b) show the second-order nonlinear OR coefficient $\chi_{0}^{(2)}$ as a function of incident photon energy $\hbar\omega$ for three different $\alpha$ with $\sigma=0.02$ and for three different $\sigma$ with $\alpha=1.2\times10^{8}radm^{-1}$, respectively.
From Fig. 4(a), it can be clearly seen that for each $\alpha$, the $\chi_{0}^{(2)}$ as a function of $\hbar\omega$ has a maximum value which occurs because of the one-photon resonance enhancement. Moreover, it is readily observed that with the increase of $\alpha$, the peak value of $\chi_{0}^{(2)}$ quickly decreases and the position of the peak value moves to left side, which shows a red shift. This behaviors can be explained as follows: with the increase of $\alpha$, the quantum-confinement of the electron reduces quickly. Therefore, the energy intervals become narrower, and as a result, the peak of the $\chi_{0}^{(2)}$ suffers a redshift.
On the other hand, Fig. 4(b) also shows that the $\chi_{0}^{(2)}$ as a function of $\hbar\omega$ has a maximum value with the same physical origin, but it is easily found that the peak position of $\chi_{0}^{(2)}$ moves to right side, which shows the blue shift, due to the augment of the energy difference between the initial state and the final state with increasing $\alpha$. In conclusion, the $\chi_{0}^{(2)}$ is greatly dependent on the two parameters for a reason that the asymmetry of the hyperbolic potential QD is strongly affected by them.

For the purpose of better understanding the relationship between nonlinear OR coefficient and the two parameters, it is useful for us to study the dependence of the geometric factor on the parameters of the hyperbolic potential. In Fig. 5, the geometric factor $\mu_{12}^2\delta_{12}$ is displayed as a function of $\alpha$ for three different $\sigma$. From which, one can find that
the value of the geometric factor $\mu_{12}^2\delta_{12}$ monotonically
decreases as both $\alpha$ and $\sigma$ increase. This is because the geometrical factor depends on the confinement strength of the hyperbolic potential QW, and the increasing $\alpha$ leads to an obvious reduction of the confinement strength, which result in a decrement of the overlap integral $\mu_{12}$ and the mean electron displacement $\delta_{12}$, respectively. For $\sigma$, the consequences are similar. As a conclusion, the geometric factor and the corresponding peak of $\chi_{0}^{(2)}$ are strong dependence of $\alpha$ and $\sigma$ in the hyperbolic QW.

In Fig. 6, the nonlinear optical absorption (OA) is presented as a function of $\alpha$ for three different $\sigma$. From this figure, one can clearly note that the nonlinear OA decreases rapidly as $\alpha$ increases, and this physical behavior can be explained as follows: the increasing $\alpha$ causes an enhancement of the electron geometric confinement of electron, thus, the electron becomes more energetic, so it penetrates easily into the potential, and the penetration can lead to a reduction of overlap between the $\psi_{0}$ state and the $\psi_{1}$ state. As a result, the overlap integral $\mu_{12}$ weakens. In addition, the increasing $\alpha$ can also lead to the fact that the energy levels become closer to each other, so the energy difference reduces with the increment of $\alpha$.
Furthermore, we can see that the nonlinear OA is more sensitive to $\sigma$ in comparison with $\alpha$. For instance, the magnitude of OA increases very quickly as $\sigma$ increases. The physical origin is that rapid enhancements of the energy interval and the geometric factor are occurred because of the variation in $\sigma$. So by tuning the values of the $\sigma$ and $\alpha$, one can get some special consequences needed in the experiment, which could be very significant in designing devices.

\section{\textbf{Summary}}

A detailed study of the optical properties have been presented for a typical $GaAs/AlGaAs$ QW with hyperbolic confinement potential. We have investigated the effects of $\alpha$ and $\sigma$ on the transition energy, oscillator strength, nonlinear OR and nonlinear OA. The energy levels and the corresponding wavefunctions are obtained within the framework of an effective mass approach. As indicated in results, an increment of $\alpha$ reduces the transition energy, oscillator strength, geometric factor, nonlinear OR and nonlinear OA. However, an increment of $\sigma$ enhances these physical quantities, but not for the geometric factor. So one can control these physical quantities by modulating the values of these parameters to meet some special needs in designing electro-optical devices.
In conclusion, optical properties in a QW with hyperbolic confinement potential
are strong dependence of $\alpha$ and $\sigma$, and we hope this theoretical study can stimulate experimental studies and practical applications.

\section{\textbf{Acknowledgement}}
This work is supported by National Natural Science
Foundation of China (under Grant No. 11074055).

\newpage

\section{caption}

Fig. 1. The shape of the hyperbolic potential well for the different $\alpha$ with $\sigma=0.02$ and the different $\sigma$ with $\alpha=2.0\times10^{8}radm^{-1}$, respectively.\\

Fig. 2. The transition  energy in a hyperbolic potential QW as a
function of $\alpha$ for three different values of $\sigma$.\\

Fig. 3. The variation of the oscillator strength as a function of $\alpha$ for different $\sigma$.\\

Fig. 4. The second-order nonlinear OR coefficient $\chi_{0}^{(2)}$ as a function of incident photon energy $\hbar\omega$ for three different $\alpha$ with $\sigma=0.02$ and for three different $\sigma$ with $\alpha=1.2\times10^{8}radm^{-1}$, respectively.\\

Fig. 5. The geometric factor $\mu_{12}^2\delta_{12}$ is displayed as a function of $\alpha$ for three different $\sigma$.\\

Fig. 6. The nonlinear OA is presented as a function of $\alpha$ for three different $\sigma$.\\

\newpage
\begin{figure}[tbp]
\begin{center}
\includegraphics[scale=1.0]{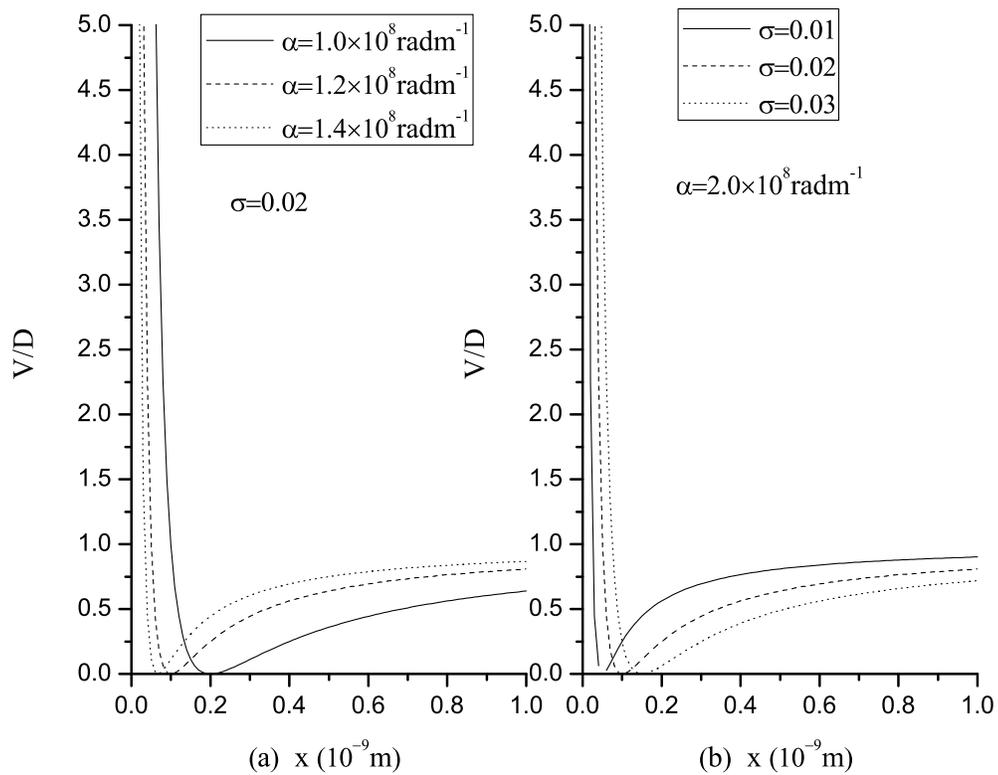}
\end{center}
\caption{The shape of the hyperbolic potential well for the different parameter $\alpha$ with $\sigma=0.02$ and the different parameter $\sigma$ with $\alpha=2.0\times10^{8}radm^{-1}$, respectively.}
\end{figure}

\begin{figure}[tbp]
\begin{center}
\includegraphics[scale=1.0]{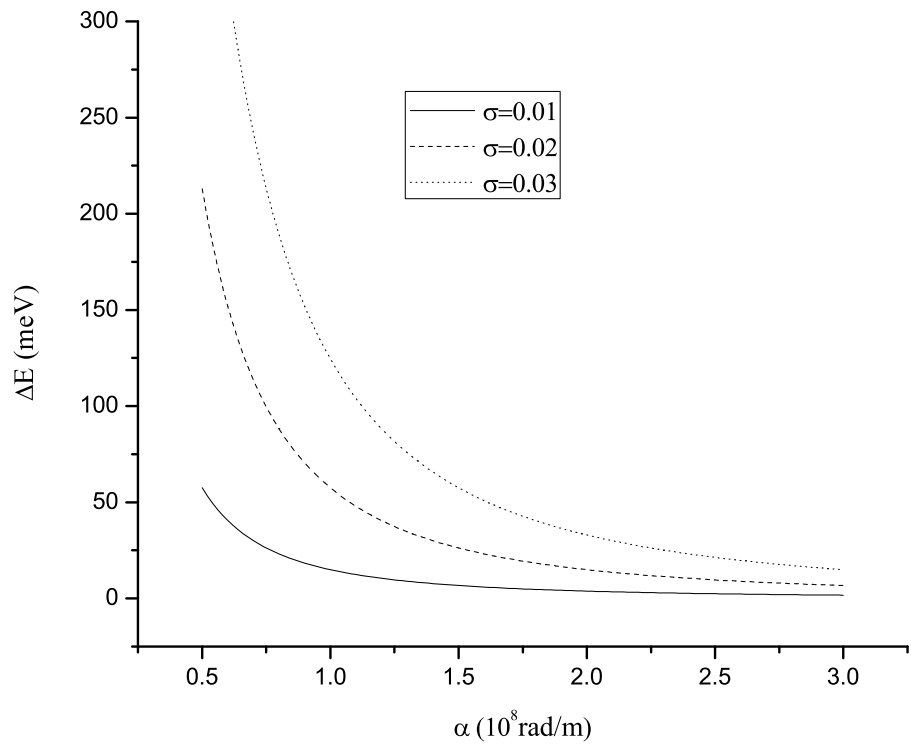}
\end{center}
\caption{The transition  energy in a hyperbolic potential QW as a
function of parameter $\alpha$ for three different values of $\sigma$.}
\end{figure}

\begin{figure}[tbp]
\begin{center}
\includegraphics[scale=1.0]{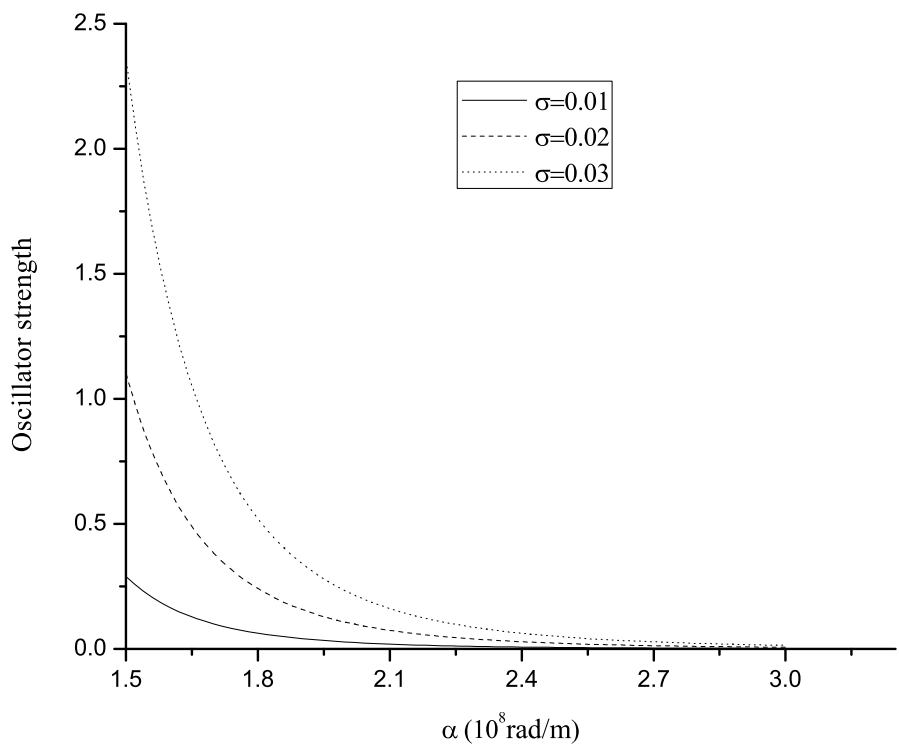}
\end{center}
\caption{The variation of the oscillator strength as a function of $\alpha$ for different $\sigma$.}
\end{figure}

\begin{figure}[tbp]
\begin{center}
\includegraphics[scale=1.0]{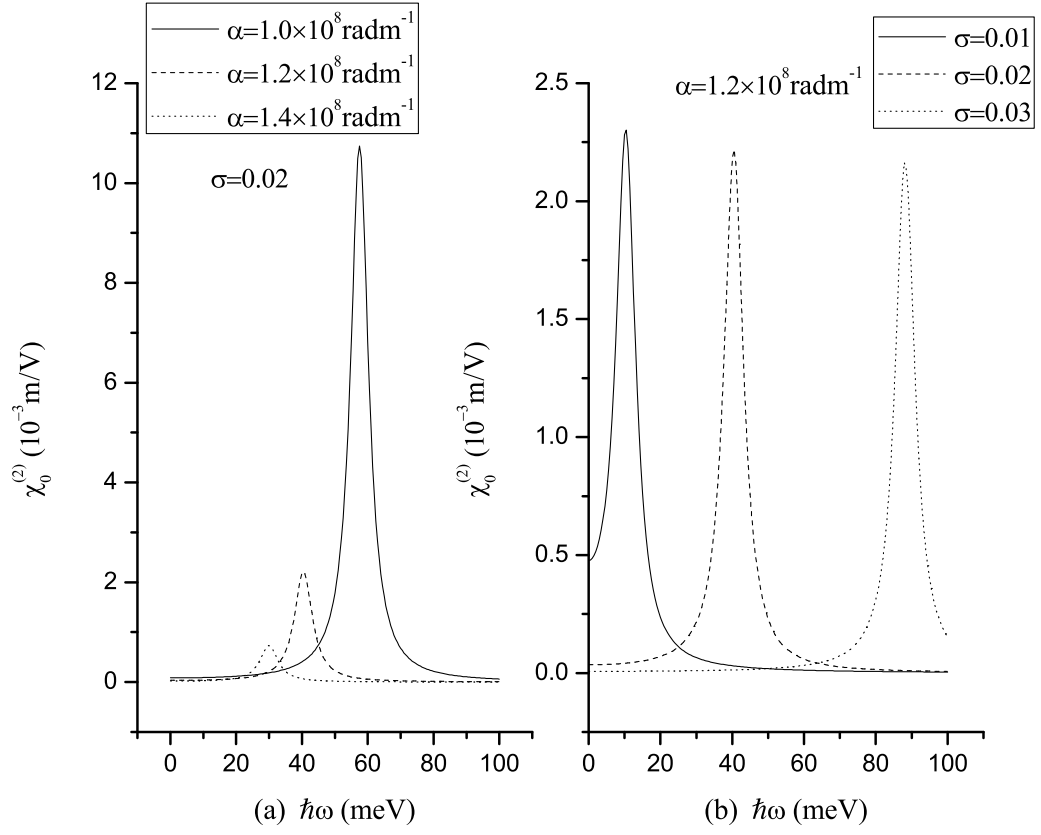}
\end{center}
\caption{The second-order nonlinear OR coefficient $\chi_{0}^{(2)}$ as a function of incident photon energy $\hbar\omega$ for three different $\alpha$ with $\sigma=0.02$ and for three different $\sigma$ with $\alpha=1.2\times10^{8}radm^{-1}$, respectively.}
\end{figure}

\begin{figure}[tbp]
\begin{center}
\includegraphics[scale=1.0]{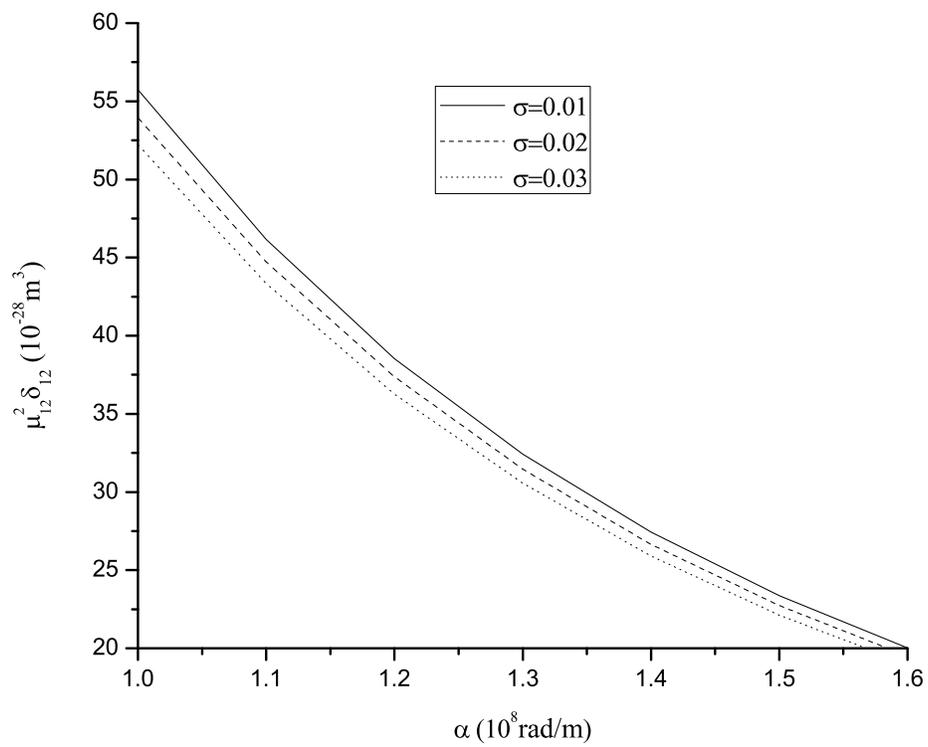}
\end{center}
\caption{The geometric factor $\mu_{12}^2\delta_{12}$ is displayed as a function of $\alpha$ for three different $\sigma$.}
\end{figure}

\begin{figure}[tbp]
\begin{center}
\includegraphics[scale=1.0]{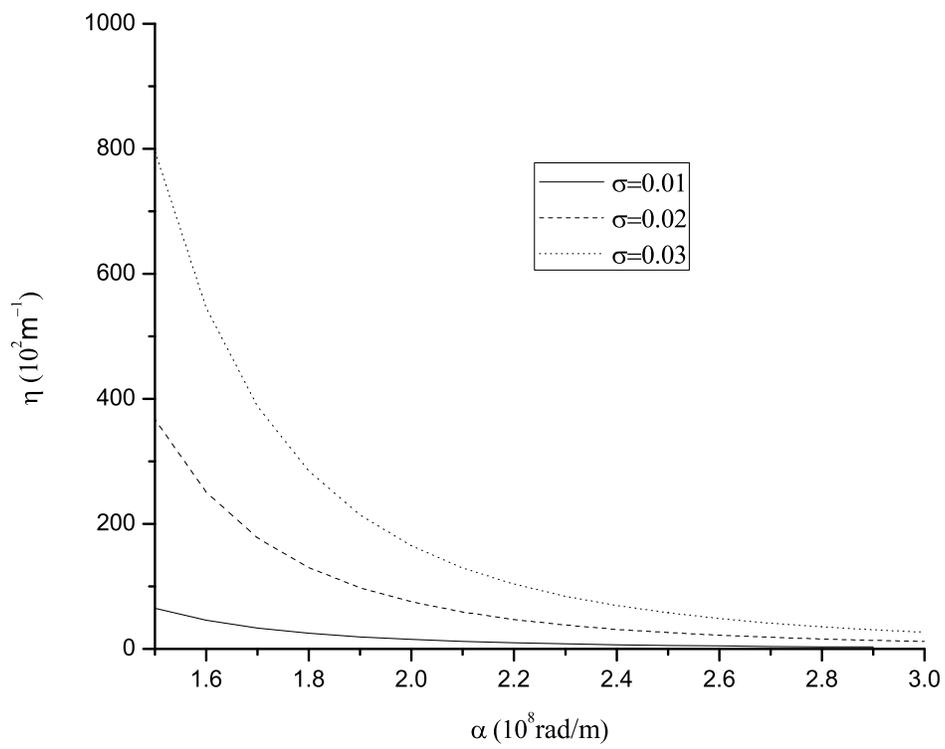}
\end{center}
\caption{The nonlinear OA is presented as a function of $\alpha$ for three different $\sigma$.}
\end{figure}

\end{document}